\documentclass[pra,letterpaper,aps,10pt,superscriptaddress,twocolumn,floatfix,showpacs]{revtex4-1}
\usepackage{graphicx}
\usepackage{amsmath}
\usepackage{amsfonts}
\usepackage{float}
\usepackage{amssymb}
\usepackage{epsfig}
\usepackage{epstopdf}
\usepackage{bbold}
\DeclareGraphicsExtensions{.pdf,.eps,.png,.jpg,.mps}
\usepackage[pdftex]{color}
\usepackage{amsmath,graphicx,amssymb,braket,xcolor,subfigure,upgreek}
\usepackage[colorlinks, linkcolor=blue, citecolor=blue, urlcolor=blue, breaklinks=true]{hyperref}
\usepackage{microtype}
\usepackage{bbm}
\usepackage{color}

\begin{document}

\title{Excitation transport with collective radiative decay}
\author{Francesca Mineo and Claudiu Genes}
\affiliation{Max Planck Institute for the Science of Light, Staudtstra{\ss}e 2,
D-91058 Erlangen, Germany}
\affiliation{Department of Physics, University of Erlangen-Nuremberg, Staudtstra{\ss}e 2,
	D-91058 Erlangen, Germany}
\date{\today}

\begin{abstract}
We investigate a one-dimensional quantum emitter chain where transport of excitations and correlations takes place via nearest neighbor, dipole-dipole interactions. In the presence of collective radiative emission, we show that a phase imprinting wavepacket initialization procedure can lead to subradiant transport and can preserve quantum correlations. In the context of cavity mediated transport, where emitters are coupled to a common delocalized optical mode, we analyze the effect of frequency disorder and nonidentical photon-emitter couplings on excitation transport.
\end{abstract}

\maketitle
\section{Introduction}
The transport of excitations, energy, charge or correlations is a topic of current interest both in the classical as well as in the quantum regimes. For example, efficient and coherent transport of excitations has been shown to play a crucial role in biological processes such as photosynthesis~\cite{engel2007evidence, scholes2011lessons, lee2007coherence}, which has inspired proposals for improvement of light collection and harvesting in solar cells~\cite{menke2013tailored}. In realistic scenarios, disorder and imperfections lead to an inhibition of transport, rendering it necessary to design strategies to combat such detrimental effects ~\cite{anderson1958absence, segev2013anderson, akselrod2014visualization, chabanov2000statistical, dalichaouch1991microwave, lahini2008anderson, lahini2009observation, john1984electromagnetic, john1987strong, schwartz2007transport, wiersma1997localization}. A simple toy model for testing possible scenarios where disorder can be circumvented is a one-dimensional chain of two-level systems: here,  in the single excitation subspace comparisons of analytical results with large scale numerics are possible. The excitation hopping can be included as stemming from the vacuum-induced dipole-dipole coupling seen as an exchange interaction. Diagonal, or frequency disorder can be included as a natural consequence of inhomogeneous broadening, as different sites see different local environments leading to an imprecision in the definition of each site's natural transition frequency. Non-diagonal, or tunneling, disorder comes from the random positioning of the sites and therefore by a varying strength of the dipole-dipole interactions between nearest neighbors.\\
\begin{figure}[b]
\includegraphics[width=0.80\columnwidth]{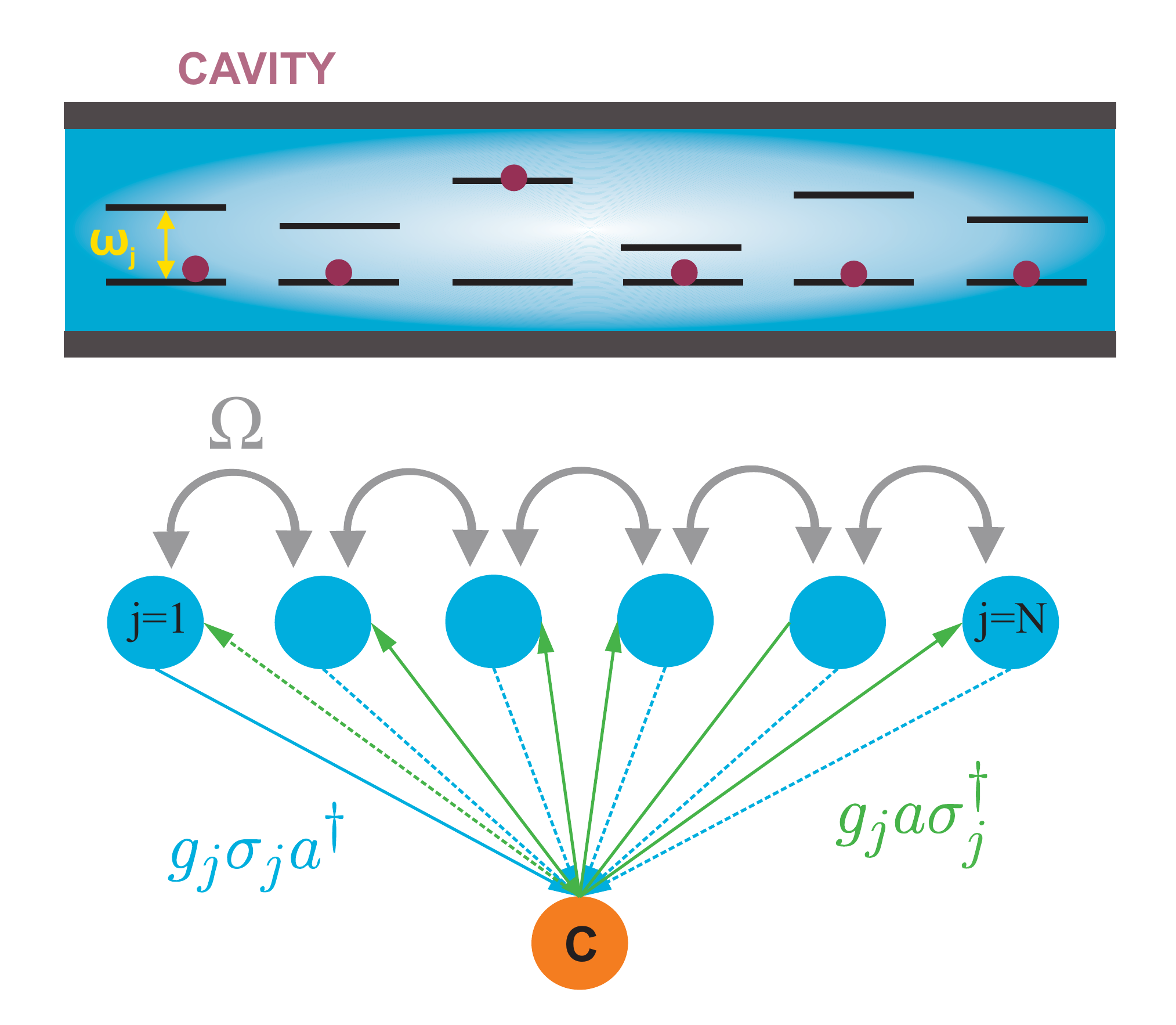}
\caption{Sketch of a frequency disordered chain of two-level quantum systems inside an optical cavity (top). Schematics of the interactions in the system where the cavity works as a \textit{bus} mode providing an additional channel of long-range transport which can overcome the slower dipole-dipole mediated mechanism (bottom).}
\label{fig1}
\end{figure}
\indent In the context of the simplified one-dimensional model treated here (illustrated in Fig.~\ref{fig1}) it has been shown~\cite{schachenmayer2015cavity} that, in the strong-coupling regime of light-matter platforms, the common coupling of $N$ sites to a single delocalized optical cavity mode can provide a scaling of transport inhibition from exponential to $N^{-2}$. This can be seen as a collective effect where the coupling of all sites to a common polaritonic 'bus' cavity mode~\cite{butte2006room, coles2014polariton, houdre1994measurement, hutchison2012modifying, kasprzak2006bose, kena2010room, lidzey1998strong, nelson1996room, weisbuch1992observation} leads to long-range interactions surpassing the efficiency of the nearest-neighbor excitation hopping process~\cite{feist2015extraordinary, zhong2016non, zhong2017energy, biondi2014self,reitz2018energy}. A different kind of collective delocalized states are encountered in densely packed ensembles of emitters where the common coupling to the infinite number of electromagnetic vacuum modes leads to superradiant/subradiant quantum superpositions~\cite{dicke1954coherence,ficek1987quantum,ficek2002entangled} exhibiting larger/smaller radiative loss than an individual, isolated two-level system. This mechanism can provide protection of excitations against decay~\cite{needham2019subradiance}. Efficient targeting of subradiant collective states has also been shown via tailored pumping where a sequence of phases are imprinted on a chain of coupled quantum emitters or via a combination of laser pulses and magnetic field gradients~\cite{plankensteiner2015selective}.\\
\indent We analyze possibilities of providing robustness of transport with respect to radiative loss in free space and to diagonal disorder in a cavity setting. In the free space scenario, we provide a partially analytical approach to the question of transport in the presence of decay and describe a phase imprinting mechanism for accessing asymmetric collective subradiant states with minimal radiative loss. Moreover, we analytically and numerically describe the preservation of quantum correlations between two propagating excitations, which we quantify by their concurrence as a measure of entanglement. In the cavity setting, we extend results from Ref.~\cite{schachenmayer2015cavity} to provide conditions for polaritonic transport with asymmetric cavity coupling in the presence of diagonal, frequency disorder.\\
\indent Section~\ref{Model and equations} introduces a simplified model of interacting quantum emitters coupled to a cavity mode and undergoing collective decay in the single excitation regime. Section~\ref{Free space transport} analytically and numerically describes the initialization and diffusion/propagation of a Gaussian wavepacket on a subradiant array and quantifies the robustness of quantum correlations between two propagating wavepackets. Section~\ref{Cavity transport} provides analytical results for polariton-mediated transport with asymmetric cavity coupling and numerical simulations for diagonal disorder.\\

\section{Model and equations}
\label{Model and equations}
We consider a chain of two-level systems (TLS) positioned at $\mathbf{r}_j$ with ground and excited states $\ket{g}_j$ and $\ket{e}_j$ for $j=1,S$. In some cases we will take $S=N$ where $N$ is the number of emitters within an optical cavity volume while in some particular cases we will consider $S=N+2M$ where in-coupling and out-coupling chains of $M$ emitters are added. The second case is useful in treating the problem of resonantly passing an excitation wavepacket through the cavity. Moving from one case to another simply requires setting $M=0$. We first write the master equation for the system which can be used either to derive equations of motion for averages which we will denote as a couple-dipoles model or to reduce the dynamics to the single excitation subspace which we dub as the quantum model.
\subsection{Master equation} \label{master_equation}
The free Hamiltonian of the system is written in terms of ladder operators $\sigma_j=\ket{g}_j\bra{e}_j$ and $\sigma_j^\dagger$ as $\mathcal{H}_0= \sum_{j} \omega_{j}\sigma^\dagger_j  \sigma_j$ (notice that we set $\hbar=1$ and the Hamiltonian could be reexpressed in terms of population inversion operators $\sigma^z_j=2\sigma^\dagger_j  \sigma_j-1$). Diagonal disorder can be included by assuming a given frequency distribution $\omega_j=\omega+\delta_j$ where $\delta_j$ is some zero-averaged distribution. The emitters see the same vacuum electromagnetic modes which give rise, after elimination, to dipole-dipole interactions of magnitude $\Omega_{ij}=\Omega(|\mathbf{r_{ij}}|)$, with $\mathbf{r_{ij}}=\mathbf{r_{i}}-\mathbf{r_{j}}$. The dipole-dipole contribution yields $\mathcal{H}_\Omega= \sum_{j \neq i} \Omega_{ij} \sigma^\dagger_{j} \sigma_{i}$, where $\Omega_{ij}$ strongly depends on the interparticle separation $r_{ij}$, the angle of the transition dipole with respect to the interparticle axis $\theta$ and the single particle independent decay rate $\gamma$ (see Appendix~\ref{A}). As in the near field the dipole-dipole interaction scales as $1/r_{ij}^3$ one can typically make the nearest neighbor approximation, therefore considering that the only non-vanishing coupling strengths are given by $\Omega_{j, j+1}=\Omega$. The TLS can be placed within the delocalized mode of an optical cavity of frequency $\omega_c$ and bosonic annihilation operator $a$, modeled by the Tavis-Cummings Hamiltonian
\begin{equation}
\mathcal{H}_c=  \omega_c a^{\dagger} a+ \sum_{j} g_{j} (a^\dagger \sigma_j + a \sigma_j^\dagger)
\end{equation}
where $g_{j}$ is the coupling between the emitter $j$ and the cavity. Collective radiative decay is included in Lindblad form ${\cal{L}}_{\text{rad}}[\rho]=\sum_{jj'} \gamma_{jj'} [\sigma_{j} \rho \sigma^{\dagger}_{j'}- (\sigma_{j}^\dagger \sigma_{j'}\rho +\rho\sigma_{j}^\dagger \sigma_{j'})/2]$. The matrix $\gamma_{ij}$ describes both independent and mutual decay processes. Notice that $\gamma_{ij}$ strongly depends on the same parameters as $\Omega_{ij}$ as they both stem from the same physical mechanism (see Appendix~\ref{A}). The cavity photon loss is described by ${\cal{L}}_{\text{cav}}[\rho]= \kappa[a \rho a^{\dagger}-(a^\dagger a\rho +\rho a^\dagger a)/2]$. With the total Linblad term ${\cal{L}}[\rho]={\cal{L}}_{\text{rad}}[\rho]+{\cal{L}}_{\text{cav}}[\rho]$, the dynamics of the system can then be followed by solving the open system master equation
\begin{equation}
\dot{\rho}(t)=-i\left[\mathcal{H},\rho\right]+{\cal{L}}[\rho],
\end{equation}
where $\rho$ refers to both emitter and cavity states. From the master equation we can derive a set of coupled equations of motion for the averages $\alpha=\braket{a}$ and $\beta_i=\braket{\sigma_i}$. The equations can be linearized in the limit of weak excitation where $\braket{\sigma_j^\dagger \sigma_j}\ll 1$ (average population of each emitter is much smaller than unity) to lead to
\begin{subequations}
\begin{align}
\label{CoupledDipoles}
\dot{\beta}_i &= -(\frac{\gamma_i}{2}+i\omega_i) \beta_i-i g_i \alpha-\sum_{j} (i\Omega_{ij}+\frac{\gamma_{ij}}{2})\beta_j,\\
\dot{\alpha} &= -(\frac{\kappa}{2}+i\omega_{\text{c}})\alpha-i\sum_{j}g_j\beta_j.
\end{align}
\end{subequations}
We will refer to this formulation as the coupled dipole model as in the weak excitation regime the dynamics is equivalent to that of a coherently and incoherently coupled system of oscillators.
\subsection{The single excitation approximation}
We construct the ground state as $\ket{G}=\ket{g_1,...g_S0_\text{ph}}$ with all spins down and no cavity photons and excited states as $\ket{j}=\ket{g_1,...e_j,...g_S0_\text{ph}}$ for $j=1,...,S$ and $\ket{S+1}=\ket{g_1,...g_j,...g_S1_\text{ph}}$ for the excitation residing inside the cavity mode. In consequence, when restricting the dynamics to a single excitation, the master equation requires the solution for $(S+2)\times(S+2)$ elements. Similarly to the approach of Ref.~\cite{needham2019subradiance} (but with an extension to include the cavity photon state as well as disordered frequencies) we derive simplified equations of motion that sees the ground state and the excited state manifold decoupling:
\begin{subequations}
\begin{align}
\dot{\rho}_{GG} &= \sum_{i,j}\gamma_{ij} \rho_{ij}\\
\dot{\rho}_{Gj} &= i\omega_j \rho_{Gj} + i \sum_{k}\rho_{Gk} \left[\Omega_{kj}+\frac{i}{2}\gamma_{kj}+G_{kj}\right]\\
\dot{\rho}_{ij} &= -i \sum_{k} \rho_{kj} \left[\Omega_{ik}-\frac{i}{2}\gamma_{ik}+G_{ik}+\omega_i\delta_{ik}\right]\\ \nonumber
                &+  i \sum_{k} \rho_{ik} \left[\Omega_{kj}+\frac{i}{2}\gamma_{kj}+G_{kj}+\omega_j\delta_{kj}\right].
\end{align}
\end{subequations}
with the cavity-coupling being $G_{ij}=g_i \delta_{j,S+1}+g_j \delta_{i,S+1}$.
One can now simply follow the evolution of the reduced density matrix in the single excitation manifold and write $\dot{\rho}_E=-i(Z\rho_E-\rho_E Z^*)$, where $Z_{ij}=\Omega_{ij}-\frac{i}{2}\gamma_{ij}+G_{ij}+\omega_j \delta_{ij}$. A quantity that one can numerically follow is the cavity transmission function~\cite{schachenmayer2015cavity} $T (t) = \sum_{j=M+N+1}^S \rho_{jj}(t)$ that quantifies the amount of excitation found on the out-coupling island.

\section{Free space transport}
\label{Free space transport}
Before moving on to analyze the effect of a delocalized bosonic cavity field we aim at elucidating a few aspects of transport in free space when collective super- and subradiant states are taken into account. We will mainly consider the coupled dipoles model where we initialize a propagating wavepacket containing on average less than one excitation. The initialization stage could be done for example by applying a short pulse from a laser with a Gaussian profile and with a propagation direction tilted with respect to the chain axis. We describe diffusion and propagation with independent decay after which we show how subradiance can provide a protection of the excitation. We then analyze, within the quantum model, the propagation of two initially entangled wavepackets where quantum correlations are quantified by concurrence as a measure of entanglement.
\begin{figure}[b]
  \centering
    \includegraphics[width=0.90\columnwidth]{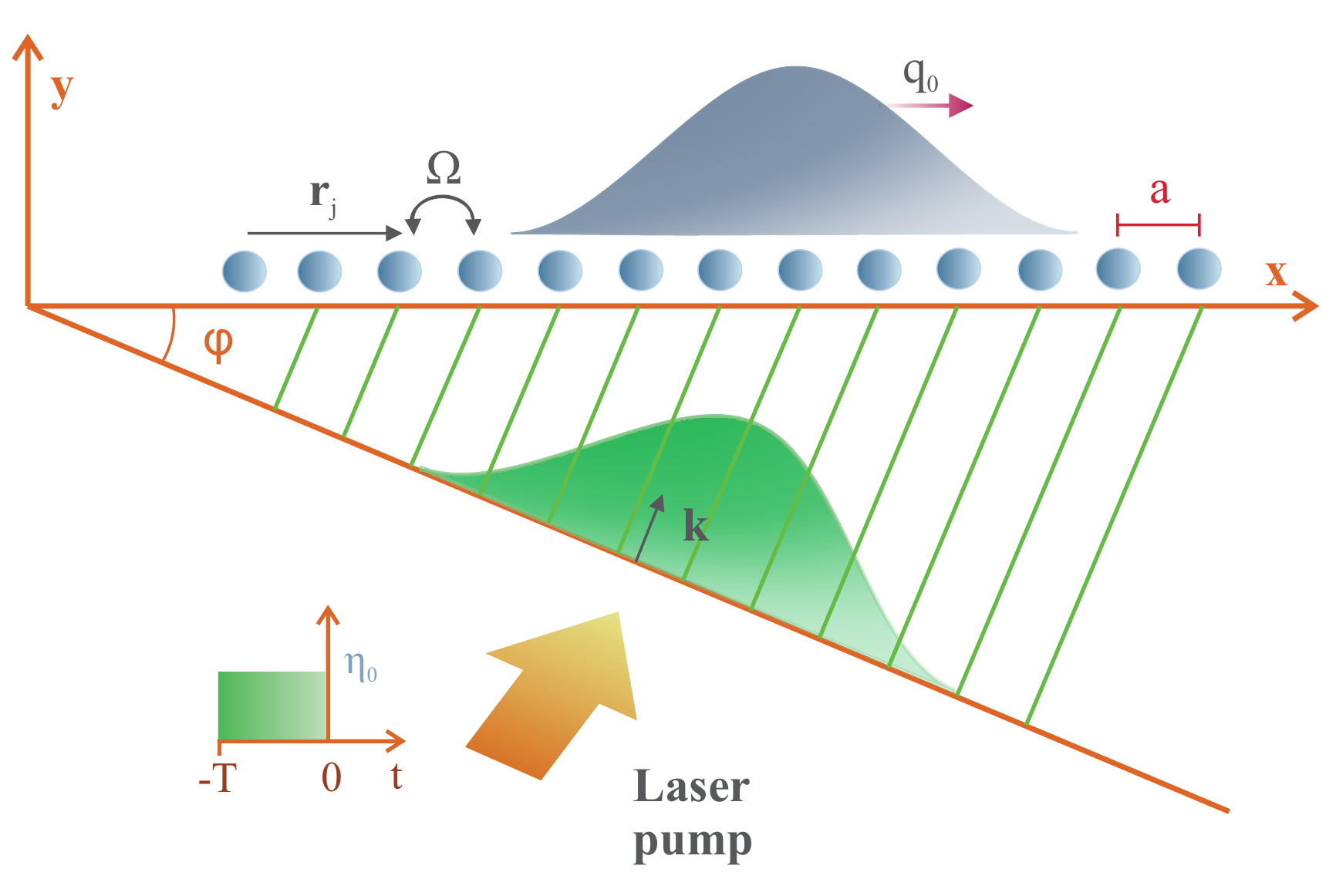}
		\caption{Initialization scheme for the Gaussian wavepacket of excitation on a chain of near-field coupled emitters, achieved by a short laser pulse of duration $T$. For $t>0$, the imprinted excitation will propagate to the right with a quasimomentum $q_0$.}
		\label{fig2}
\end{figure}
\subsection{Wavepacket evolution with independent decay}
We initialize a Gaussian wavepacket providing a weak excitation onto the system via an external tilted laser field with a Gaussian profile in amplitude (as depicted in Fig.~\ref{fig2}). The driving Hamiltonian reads
\begin{equation}
\mathcal{H}_{\text{drive}}=\sum_{j} \eta_{j}(t) \left[\sigma_{j} e^{i \omega_\ell t} e^{i \mathbf{k} \cdot  \mathbf{r_{j}}}+\text{h.c.}\right],
\end{equation}
where $\omega_\ell$ is the laser frequency and $\mathbf{r}_j= a j \mathbf{x}$ describes positioning within an equidistant chain in the x-direction with lattice constant $a$. Notice that the tilting of the laser is equivalent to imprinting a quasi-momentum $q_0=k a \sin\phi$ derived from $\mathbf{k} \cdot \mathbf{r_{j}}=(k \sin\phi)(j a)=(k a \sin\phi)j=q_0 j$. The pulse is assumed constant between $t=-T$ and $t=0$ and the excitation amplitude follows a Gaussian profile with $f_{j}=1/\sqrt{\sqrt{2 \pi}w}e^{-(j-j_0)^2/(4w^2)}$. We assume that the pulse is fast enough ($T<\Omega^{-1}$) such that no hopping of excitations can occur during the driving. This allows one to neglect the dipole-dipole interaction during the initialization stage and derive a simple equation of motion for the coherences at each site:
\begin{equation}
\partial_t{\braket{\sigma_j}} =-i\omega_j \braket{\sigma_j}+ i \eta_j \braket{\sigma_j^z} e^{-i q_0 j} e^{-i \omega_\ell t}.
\end{equation}
Within the low-excitation approximation obtained by assuming that $\braket{\sigma_j^z} \sim -1$ and making the following notation $\beta_j=\braket{\sigma_{j}}$ one can rewrite the equation of motion for the $j^{\text{th}}$ dipole moment in a frame rotating at the laser frequency
\begin{equation}
\dot{\beta_j}=-i\Delta_j \beta_j-i \eta_{j}(t) e^{i q_0 j},
\end{equation}
where $\Delta_j=\omega_j-\omega_\ell$. Since the equations are decoupled we can integrate them for the time of the pulse $-T<t<0$ and with initial condition $\beta_j(-T)=0$ (no excitation before the pump) to obtain
\begin{equation}
\beta_j(0) =2 i \eta_0 T f_j  \frac{\sin\left(\Delta_j T/2\right)}{\Delta_j T/2}e^{-i q_0 j}=\beta_0 e^{-i q_0 j} f_j.
\end{equation}
To insure that the weak excitation condition is fulfilled we will impose the condition that the total population in the chain (and in the absence of disorder such that $\Delta_j=\Delta$) under resonance conditions $\sum_j |\beta_j(0)|^2=4(\eta_0 T)^2$ is much less than unity.\\
\indent After the initialization stage, we follow the evolution of the wavepacket for $t>0$ in the presence of hopping under the Hamiltonian $\mathcal{H}_{t>0}=\mathcal{H}_0+\mathcal{H}_\Omega$ and diagonal independent decay. To this purpose we write Eqs.~\ref{CoupledDipoles} (in the absence of the cavity mode and assuming all hopping rates equal to $\Omega$, all decay rates equal to $\gamma$ and all frequencies $\omega$) in a general form $\dot{\vec{\beta}}=- M \vec{\beta}$ where
\begin{align}
M_{jj'}=(i\omega+\gamma/2)\delta_{jj'}+i\Omega (\delta_{j,j'+1}+\delta_{j,j'-1}).
\label{eqM}
\end{align}
We have already assumed that the dipole-dipole exchange can be reduced to a nearest-neighbor interaction and that we are in the case of open boundary condition (OBC). For periodic boundary conditions (PBC) we would add two extra terms $i\Omega (\delta_{j,1}\delta_{j',S}+\delta_{j,S}\delta_{j',1})$ which couple the first with the last emitter in the chain.\\
\indent Notice that the evolution matrix can be diagonalized by the same transformation that diagonalizes the Toeplitz matrix such that one can write $M=V\Lambda V^{-1}$. Assuming PBC we have
\begin{align}
\Lambda_{kk'}=i[\omega+2\Omega \cos{(k\theta)}-i\gamma/2]\delta_{kk'}=(i\mathcal{E}_k+\gamma/2)\delta_{kk'},
\end{align}
(with $\theta=2\pi/S$) and the matrix of eigenvectors has the following elements $V_{jk}=e^{-i  j k\theta }/\sqrt{S}$. Notice that this matrix is symmetric as $V_{jk}=V_{kj}$ and for the inverse matrix we have $[V^{-1}]_{jk}=V^*_{jk}$. Here, the index $k$ run from $0$ to $S-1$ while the index $j$ runs from $1$ to $S$. For OBC the eigenvalues are unchanged but one redefines $\theta=\pi/(S+1)$ and obtains real eigenvectors $V_{jk}=\sqrt{2/(S+1)}\sin{(\theta j k)}$ with the same properties as for PBC and all indexes run from $1$ to $S$.\\
\indent We can now generally write the solution for all dipole amplitudes as $\vec{\beta}(t) =  V e^{-\Lambda t} V^{-1}  \vec{\beta}(0)$. More explicitly, for each component:
\begin{equation}
\beta_j(t) = \beta_0 \sum_{k,j'} e^{-i\mathcal{E}_k t-i q_0 j'}e^{-\gamma t/2}V_{jk} V^*_{kj'} f_{j'}.
\label{explicit-time-evolution}
\end{equation}
The sum over the initial Gaussian distribution of excitation can be analytically estimated in the particular case that the wavepacket is not too narrow. In the Fourier domain, this means that we ask for the $k$ distribution around the central value $k_0=q_0/\theta$ to be small such that a Taylor expansion of the energy dispersion relation is possible (see Fig.~\ref{fig3}a):
\begin{equation}
\mathcal{E}_k \simeq \mathcal{E}_{k_0}-2 \Omega \theta \sin{(k_0 \theta)} (k-k_0)- \Omega \theta^2 \cos{(k_0 \theta)}(k-k_0)^2.
\label{eq-Omega-k}
\end{equation}
In the general case, under the approximation that a second order Taylor expansion suffices, the wavepacket maintains a Gaussian character and we can analytically describe the distribution of excitation in time as
\begin{equation}
|\beta_j(t)|^2= |\beta_0|^2 \frac{1}{\sqrt{2 \pi} \bar{w}(t)} e^{-\frac{\left[j-\bar{j}(t)\right]^2}{2 \bar{w}(t)^2}}e^{-\gamma t}.
\label{p-nodecay}
\end{equation}
Both the wavepacket central position and its diffusion acquire a time dependence analytically expressed as
\begin{subequations}
	\begin{align}
\bar{w}^2(t) &=w^2\left[1+\frac{\Omega^2 t^2 }{w^4}\cos^2q_0\right] \label{eq:w(t)}, \\
\bar{j}(t)&=j_0+2 \Omega t \sin q_0.
\end{align}
\end{subequations}
For $0<q_0<\pi$ (for the particular choice of $\Omega>0$) the packet moves to the right, reaching the fastest speed $v_g=2\Omega \sin q_0$ at $q_0=\pi/2$, while for $\pi<q_0<2 \pi$ the packet moves to the left. Stationary diffusion is reached for $q_0=0$ or $q_0=\pi$ where $\bar{j}(t)=j_0$ and the variance increases quadratically in time at large times where $\Omega t\gg w^2$. For minimal diffusion and optimal speed one sets $q_0=\pi/2$ obtaining $\bar{w}(t)=w$ and $\bar{j}(t) = j_0+2 \Omega t$ showing the wavepacket moving with the group velocity $v_g=2 \Omega$ and unchanged in shape. Notice that in this particular case, for OBC $k_0\approx S/2$ and the energy dispersion can be approximated by a line as illustrated in see Fig.~\ref{fig3}a.\\
\indent We recall that the value of $q_0$ could be adjusted by simply varying the angle $\phi$ at the initialization such that for optimal $\phi=\pi/2$ we have $q_0=2 \pi a /\lambda$. As for considerable nearest neighbour near field coupling one needs small interparticle distances, this procedure limits the achievable values of $q_0$ to smaller than $\pi/2$ values. Therefore the achievement of these values will need an additional protocol of implementation as for example the application of a magnetic field gradient as in Ref.~\cite{plankensteiner2015selective} or a more complicated interval level scheme where particles can be trapped with fields of small wavelength while the initialization of the wavepacket could be done via a larger wavelength field.\\
\subsection{Wavepacket evolution with subradiance}
\label{subradiance free space}

\begin{figure*}[t]
\includegraphics[width=0.98\textwidth]{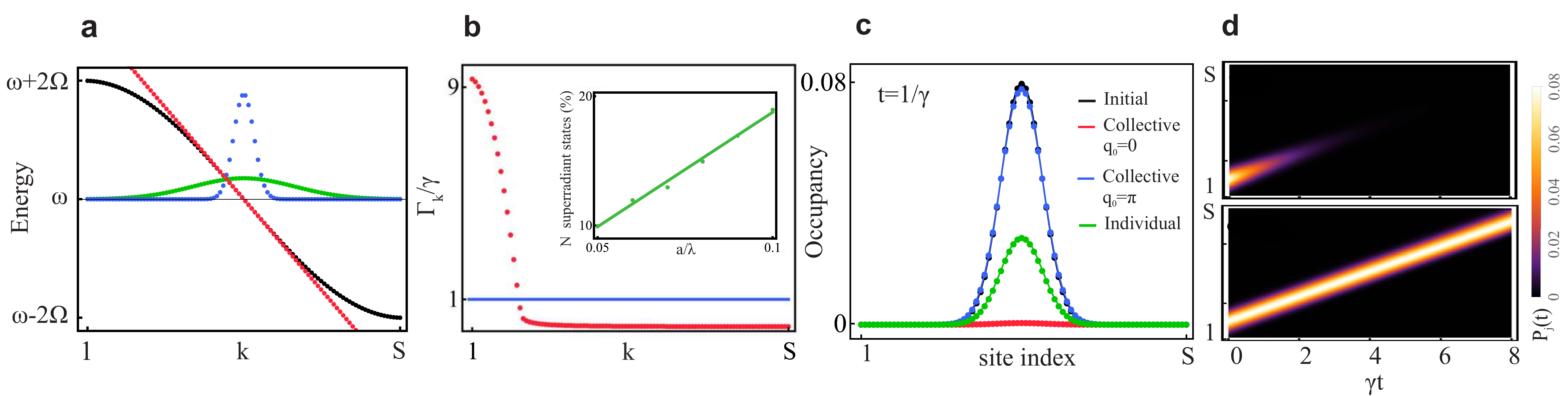}
\caption{\textbf{(a)} Energy dispersion with OBC in black ($\mathcal{E}_k$ for collective states indexed by k from $1$ to $S$). The red line shows the Taylor expansion approximation assuming $q_0=\pi/2$. The green and blue curves are the $k$-space components of two initial wavepackets with $w=1$ and $w=5$, respectively. Parameters are $S=100$ and $\Omega=0.07$. \textbf{(b)} Normalized collective decay rates $\Gamma_k/\gamma$.  The inset shows the scaling of the percentage of superradiant states ($\Gamma_k>\gamma$) with increasing interparticle separation. \textbf{(c)} Time evolution of an initial Gaussian wavepacket with independent and collective decay, where the quasimomentum  initialization allows the direct tuning into superradiant ($q_0=0$) or subradiant $(q_0=\pi)$ behaviour. The blue curve shows robustness against decay when the excitation is initially encoded in a subradiant superposition. \textbf{(d)} Time evolution of a wavepacket initialized in the left part of the chain with $q_0=\pi/2$, comparison between individual decay (top) and collective decay (bottom), considering $S=110, w=5, a/\lambda=0.08$.}
\label{fig3}
\end{figure*}

The presence of individual emitter decay has the trivial effect of exponentially reducing the excitation number during propagation. A straightforward way of tackling this detrimental aspect brought on by the radiative emission is to consider structures exhibiting robustness to decoherence, such as subradiant arrays.
For small inter-particle separations $a<\lambda$, the diagonalization of the mutual decay rates matrix $\Gamma$ gives rise to $S$ channels of decay, some of superradiant character (decay rate larger than $\gamma$) but most of them exhibiting subradiance (decay rate smaller than $\gamma$). The inclusion of the collective decay effect is done in Eq.~\eqref{eqM} by replacing $\gamma$ with $\gamma_{jj'}$. The diagonalization of the coherent part leads to $V^{-1}MV=\Lambda+V^{-1}(\Gamma/2)V$. The last will have diagonal terms $\Gamma_k=\sum_{jj'} V^*_{jk} \gamma_{jj'}V_{j'k}$ describing decay of the collective state to the ground state of the system while all non-diagonal terms describe migration of excitation within the single excitation manifold. Assuming that the diagonal parts are dominant, one can estimate that most of the collective states are subradiant, as illustrated in Fig.~\ref{fig3}b. The inset shows a roughly linear dependence of the percentage of superradiant states with decreasing interparticle separation. For small separations, where subradiant effects are strong, the number of superradiant states reduces to less that $\sim 20\%$ of the total number of states.\\
\indent Let us analyze the influence of subradiant transport in the collective basis where the collective amplitudes are defined from the transformation $\vec{\tilde{\beta}}=V^{-1} \vec{\beta}$ which on components reads $\tilde{\beta}_k=\sum_j V^*_{kj} \beta_j$. Starting from example with a single localized excitation and with OBC, the initial occupancy of each collective state is simply $1/S$. For a mesoscopic ensemble we can then estimate the survival probability of the excitation (for time $t\gg\gamma^{-1}$ after all subradiant states decayed) simply from counting the number of subradiant states in Fig.~\ref{fig3}b. For an initial Gaussian wavepacket, the occupancy of the k-th collective state is found to be also a Gaussian
\begin{equation}
|\tilde{\beta}_k|^2=|\beta_0|^2 \frac{1}{\sqrt{2 \pi} \tilde{w}_k} e^{\frac{-(k-k_0)^2}{2 \tilde{w}_k^2}}
\end{equation}
centered at $k_0 =q_0/\theta$ and with a width $\tilde{w}_k =1/(2 \theta w)=S/(4 \pi w).$\\
\indent For an initial stationary wavepacket undergoing diffusion, Fig.~\ref{fig3}c shows the impact of subradiant collective states on the preservation of the excitation. At a time $t=\gamma^{-1}$, individual decay shows the expected decrease of the wavepacket amplitude while a strategy of constant illumination (corresponding to $q_0=0$) leads to a very quick superradiant decay of the excitation. Illumination with phases of adjacent emitters alternating by $\pi$ (corresponding to $q_0=\pi$) leads instead to the immediate mapping of the collective state onto a subradiant robust one.

\subsection{Transport of correlations}
Let us now move to the alternative scenario where dynamics takes place in the single excitation Hilbert space of dimensions $S+1$ with the basis vectors made up of the collective ground state and single excitation $\ket{j}$ states. We assume that an initial entangled state is prepared as a superposition between two Gaussians centered at $j_0$ and $j_0+d_0$ where $d_0$ quantifies the distance between the two Gaussians. We recall the previous definition $f_{j}=1/\sqrt{\sqrt{2 \pi}w}e^{-(j-j_0)^2/(4w^2)}$ and define the initial state as
\begin{equation}
\ket{\psi(0)} = \frac{1}{\sqrt{2}}\sum_{j=1}^{S} e^{i q_0 j} (f_{j} +f_{j-d_0}) \ket{j}.
\end{equation}
We aim at analyzing the behavior of quantum correlations with respect to independent decay and possibly utilize the robustness brought on by collective subradiant states. To this end we employ concurrence as a measure of bipartite entanglement defined as $\mathcal{C}_{jj'}=\text{Max}\{0, \sqrt{\lambda_1}-\sqrt{\lambda_2}-\sqrt{\lambda_3}-\sqrt{\lambda_4}\},$ where the eigenvalues are computed on the matrix $\Lambda^{(jj')}= \bar{\rho}^{(jj')}(\sigma_y \otimes \sigma_y)[\bar{\rho}^{(jj')}]^* (\sigma_y \otimes \sigma_y)$ and are arranged in decreasing order. The density matrix used to compute the concurrence  is the reduced one obtained after tracing over all other particle and field states except for particles $j,j'$. As we are working in the single excitation only, the density matrix elements for double excitation are zero and the reduced matrix reads
\begin{equation}
	\bar{\rho}^{(jj')}=  \begin{bmatrix}
		\rho_{GG}+\sum_{n\neq j,j'} P_n  & \rho_{Gj}& \rho_{Gj'} & 0  \\
		(\rho_{Gj})^*      &P_j &\rho_{jj'} &0  \\
		(\rho_{Gj'})^*     &(\rho_{jj'})^* &P_{j'} &0  \\
		0      &0 &0 &0  \\
	\end{bmatrix}
\end{equation}
where $P_j=\rho_{jj}$. Notice that tracing over all particles except $j$ and $j'$ has the only consequence of increasing the weight of the zero excitation state in the reduced density matrix, while leaving all coherences (off-diagonal elements) unaffected. From here one can explicitly write the matrix $\Lambda^{(jj')}$ as
\begin{widetext}
\begin{equation}
	\Lambda^{(jj')}=  \begin{bmatrix}
		    0   & \rho_{Gj}P_{j'}+\rho_{Gj'} \bar{\rho}^*_{jj'} & \rho_{Gj} P_{j}+\rho_{Gj}\bar{\rho}_{jj'} & -2\rho_{Gj}\rho_{Gj'}  \\
		0       & P_{j}P_{j'}+|\rho_{jj'}|^2 & 2\rho_{jj} \rho_{jj'} & -\rho_{Gj'}P_{j}-\rho_{Gj}\rho_{jj'}  \\
		0       & 2P_{j'} \rho^*_{jj'} & \rho_{jj} \rho_{j'j'}+|\rho_{jj'}|^2 & -\rho_{Gj}P_{j'}-\rho_{Gj'}\rho^*_{jj'}  \\
		0       & 0 & 0 & 0  \\
	\end{bmatrix}.
\end{equation}
\end{widetext}
Surprisingly, the eigenvalues, in decreasing order, assume a very simple form independent of the coherence between the ground state and the single excitation states
\begin{equation}
\lambda_{1,2}=(\sqrt{P_j P_{j'}}\pm|\rho_{jj'}|)^2
\end{equation}
and $\lambda_{3,4}=0$. The concurrence for sites $jj'$ then can be computed from here as specified above:
\begin{equation}
\mathcal{C}_{jj'}=|\sqrt{P_j P_{j'}}+|\rho_{jj'}||-|\sqrt{P_j P_{j'}}-|\rho_{jj'}||.
\end{equation}
Notice that as decoherence mechanisms typically affect the two particle coherence rather than populations, the concurrence for two sites is simply $\mathcal{C}_{jj'}=2|\rho_{jj'}|$ and therefore easily estimated even at the analytical level. For example, between a mixed state and a Bell maximally entangled state, the concurrence varies betweeen 0 and 1 as indicated by the off diagonal elements of the density matrix. For the two non-overlapping Gaussian wavepackets, we can define an average concurrence $\mathcal{C}_\text{av}(t)=\sum_{j\in\mathcal{D}_1,j\in\mathcal{D}_2}\mathcal{C}_{jj'}/(5w(t))$ where the sum is done over the non-overlapping domains $\mathcal{D}_{1,2}$ referring to the two wavepackets. The normalization by the average number of sites participating in the entangled state gives an average concurrence close to unity. At the analytical level, it is straightforward to show that for non-diffusive, initial wavepackets made of independently decaying emitters the concurrence simply decays in time as $e^{-\gamma t}$. For collective decay, the behavior reproduces closely the one of the propagating single wavepacket: as subradiance protects both decay of population and coherence, the average concurrence stays close to unity as long as the wavepacket does not decay.

\section{Cavity transport}
\label{Cavity transport}
\begin{figure*}[t]
	\centering
	\includegraphics[width=0.98\textwidth]{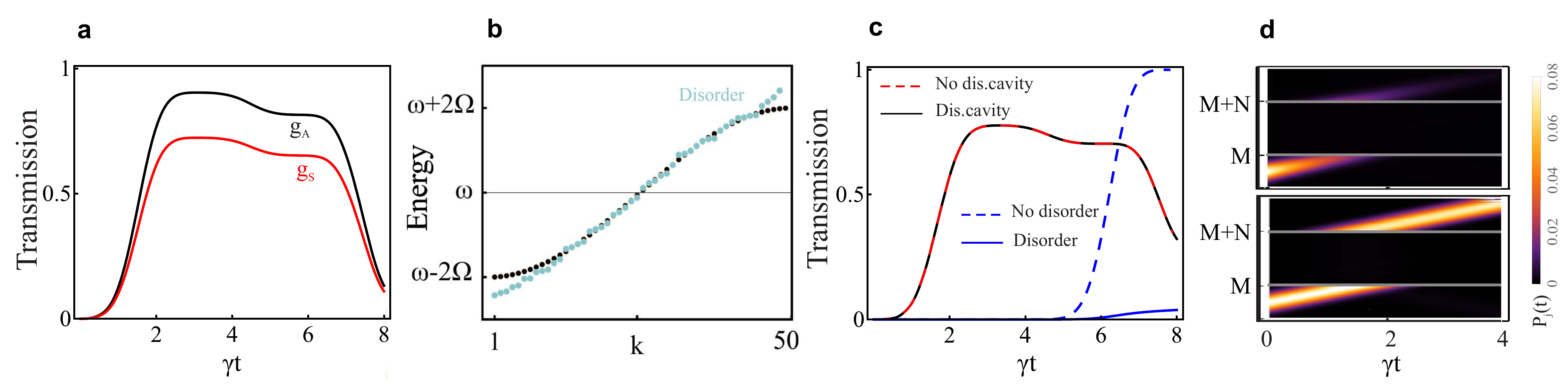}
	\caption{\textbf{(a)} Cavity transmission comparison, in the presence of collective decay, between symmetric versus asymmetric couplings scenarios with parameters $M=30, N=50, \omega_c=\omega, g= 90 \Omega, \Omega'=10\Omega, \Delta_{S,A}=\pm 635.4$. \textbf{(b)} Energy dispersion curve shows little influence from the presence of disorder (at the order of $\Omega$ for all non-polaritonic $S-1$ states shown here. \textbf{(c)} Transmission in the presence of disorder and collective decay, considering that the wavepacket is matched into the antisymmetric polariton energy in which case the cavity transport is not influenced by disorder. In contrast, free space transport is slower (dashed, blue line in the absence of disorder) and strongly inhibited by disorder (full blue line). Disorder averaging has been performed over $100$ realizations. \textbf{(d)} Time evolution of the wavepacket through cavity, considering individual decay (top) versus collective decay (bottom). The grey lines denote the cavity boundaries.}
	\label{fig4}
\end{figure*}
A way to circumvent detrimental effects of disorder in the transport of energy has been proposed in Refs.~\cite{schachenmayer2015cavity,feist2015extraordinary}. The mechanism is based on the collective coupling to a cavity delocalized mode, which leads to the occurrence of additional polariton-mediated channels for enhanced energy transport. We propose here an additional improvement by showing that when polaritons are formed by the hybridization of the photon state with asymmetric superpositions of the quantum emitters, protection of excitation can be achieved by spreading the wavepacket into robust collective subradiant states. \\
\indent In the case of identical cavity-emitter couplings $g_{j(S)}=g$, a bright mode is formed as a symmetric superposition of all quantum emitters $B=\sum_{j} \sigma_{j}/\sqrt{N}$. The corresponding bright state is obtained by applying $B^\dagger$ to the ground state $\ket{G}$ obtaining a W-state. This mode is hybridized with the cavity field leading to polaritonic states that can be obtained from the action of the following operators $p^\dagger_{u,d (S)} =1/\sqrt{2}( a^\dagger \pm \sum_{j} \sigma^\dagger_{j}/\sqrt{N})$ onto the ground state. The two light-matter hybrid quantum states are the upper (u) and lower (d) polaritonic states energetically positioned at $\omega\pm g\sqrt{N}$. Notice that the same polaritonic energies can be obtained even if the couplings follow a different symmetry scaling as for example $g_{j(A)}=(-1)^jg$, albeit with very different collective states obtained as $\sum_{j} \sigma^\dagger_{j}/\sqrt{N}\ket{G}$. As the analysis in Refs.~\cite{schachenmayer2015cavity,feist2015extraordinary} neglected collective radiative effects, the symmetry of the collective polaritonic states did not play a role. However, symmetric modes are strongly superradiant at small particle-particle separations and therefore not optimized for robust transport. A natural choice is to consider instead transport through very asymmetric, typically very subradiant states. \\
\indent Let us first consider the eigenvalue problem of the Tavis-Cummings model plus nearest-neighbour dipole-dipole exchanges. We denote the eigensystem by $\omega_n$ and $\ket{n}$ such that the eigenvalue problem becomes $\mathcal{H}\ket{n}=\omega_n\ket{n}$ for $n$ running from $1$ to $N+1$. In the single excitation regime the general form of an eigenvector will then be of the form
\begin{equation}
\ket{n}= \sum_{j=1}^N c_j^{(n)}\ket{j}+\beta^{(n)}\ket{1_\text{ph}},
\end{equation}
where normalization requires that $\sum_j |c_j^{(n)}|^2+|\beta^{(n)}|^2=1$. The task is to find all $\omega_n$ and corresponding coefficients of the emitter $c_j^{(n)}$ and photon $\beta^{(n)}$ content in each eigenvector. To this end we use the diagonal representation of the dipole-dipole interaction $\mathcal{H}_\text{dd}=2\Omega \sum_k\cos(k\theta)\ket{\tilde{k}}\bra{\tilde{k}}$ and the transformation $\ket{\tilde{k}}=\sum_{j}V_{jk}\ket{j}$ to find the representation $\mathcal{H}_\text{dd}=\sum_k\mathcal{E}_k\sum_{j}\sum_{j'}V_{jk}V^*_{j'k}\ket{j'}\bra{j}$. One can then proceed by finding a set of couple equations for $c_j^{(n)}$ and $\beta^{(n)}$ from which the eigenvalues can be extracted.\\
\indent For the symmetric case the sum $\sum_{j} V_{jk}=\sqrt{N}\delta_{k,0}$ selects only the symmetric collective mode with $k=0$ and one ends up solving for
\begin{align}
\left[(\omega_n-\omega)^2 -g^2 N\right] \beta^{(n)}-\mathcal{E}_0(\omega_n-\omega)\beta^{(n)} =0.
\end{align}
There are $N-1$ degenerate solutions with zero photonic component $\beta^{(n)}=0$ and two polariton states with energies obtained as solutions of a quadratic equation
\begin{align}
\omega_{\pm}^\text{sym}=\omega+\Omega\pm \sqrt{g^2 N+\Omega^2}.
\end{align}
For small tunneling rates $\Omega\ll g\sqrt{N}$ we can approximate the polariton energies at $\omega\pm g\sqrt{N}+\Omega$. The polaritonic states show a photon contribution
\begin{equation}
\beta^\pm=\frac{g \sqrt{N}}{\sqrt{(\omega_\pm-\omega)^2+g^2 N}},
\end{equation}
while the matter contribution is
\begin{equation}
c_j^\pm= \frac{\omega_\pm-\omega}{\sqrt{N} \sqrt{(\omega_\pm-\omega)^2+g^2 N}}.
\end{equation}
Notice that, as expected, in the absence of dipole-dipole couplings, the expressions above reduce to the expected $\beta^\pm=1/\sqrt{2}$ and $c_j^\pm=\pm1/\sqrt{2N}$.\\
\indent In the completely asymmetric case where $g_j=g(-1)^j$ we select the asymmetric mode $\sum_{j} (-1)^jV_{jk}\approx \sqrt{N}\delta_{k,N/2}$ (for PBC) and the solution is similar to that above with a slight difference in the energy of the polaritons
\begin{align}
\omega_{\pm}^\text{asym}=\omega-\Omega\pm \sqrt{g^2 N+\Omega^2}.
\end{align}
The photonic part of the asymmetric eigenvectors is identical to above while the matter contribution shows the phase dependence dictated by the coupling variation among the emitters
\begin{equation}
c_j^\pm= (-1)^j\frac{\omega_\pm-\omega}{\sqrt{N} \sqrt{(\omega_\pm-\omega)^2+g^2 N}}.
\end{equation}
Having identified the energies of the asymmetrically driven polaritons, we can compare our results with those of Ref.~\cite{schachenmayer2015cavity}. In Fig.~\ref{fig4}a transmission through a cavity with $g_j=g(-1)^j$ is shown more efficient that the overall equal coupling mechanism. In the presence of disorder, Fig.~\ref{fig4}b illustrates that the dispersion curve does not change too much. Moreover, as Ref.~\cite{sommer2020molecular} describes in detail, the polaritonic energies are also very robust with disorder even at the level of $g\sqrt{N}$. Therefore, as also concluded in~\cite{schachenmayer2015cavity} (in the case of positional disorder), diagonal disorder plays almost no role in the transmission through the cavity even if it has a strong role of localization excitations in free space, as shown in Fig.~\ref{fig4}c. Finally, Fig.~\ref{fig4}d shows robust transport in the collective radiative regime (bottom propagation line) versus independently decaying emitters.\\
\section{Conclusions}
We have treated aspects of excitation and quantum correlations propagation on a one dimensional chain of nearest neighbor coupled quantum emitters in the presence of a collective radiative bath. The robustness of collective subradiant states can be exploited towards more efficient transport of excitations by proper phase imprinting in free space. Also, not only excitations but quantum correlations as well can show robustness against radiative decay when transport takes place via subradiant collective states. In cavity settings, where a common delocalized bosonic light mode couples to all emitters, an asymmetric coupling pattern shows protection against radiative decay as well as against diagonal, frequency disorder in the chain of emitters. \\
\section{Acknowledgments}
We acknowledge financial support from the Max Planck Society and from the German Federal Ministry of Education and Research, co-funded by the European Commission (project RouTe), project number 13N14839 within the research program "Photonik Forschung Deutschland" (C.~G.). This work was also funded by the Deutsche Forschungsgemeinschaft (DFG, German Research Foundation) -- Project-ID 429529648 -- TRR 306 QuCoLiMa
(``Quantum Cooperativity of Light and Matter''). We acknowledge fruitful discussions with J.~Schachenmayer and C.~Sommer.

\bibliography{Refs}

\clearpage
\onecolumngrid
\appendix

\newpage
\section{Vacuum mediated coherent and incoherent interactions}
\label{A}

The vacuum mediated dipole-dipole interactions for an electronic transition at wavelength $\lambda$ (corresponding wave-vector $k=2\pi/\lambda$) between an identical pair of emitters separated by $r_{ij}$ is
\begin{equation}
\Omega_{ij}=\frac{3}{4}\gamma \left[(1-3\cos^2\theta)\left(\frac{\sin (kr_{ij})}{(kr_{ij})^2}+\frac{\cos (kr_{ij})}{ (kr_{ij})^3}\right)-\sin^2\theta\frac{\cos (kr_{ij})}{(kr_{ij})}\right].
\end{equation}
The quantity $\theta$ is the angle between the dipole moment $\mathbf{d}$ and the vector $\mathbf{r}_{ij}$. The associated collective decay is quantified by the following mutual decay rates
\begin{equation}
\gamma_{ij}=\frac{3}{2}\gamma \left[(1-3\cos^2\theta)\left(\frac{\cos (kr_{ij})}{(kr_{ij})^2}-\frac{\sin (kr_{ij})}{ (kr_{ij})^3}\right)+\sin^2\theta\frac{\sin (kr_{ij})}{(kr_{ij})}\right].
\end{equation}

\section{Wavepacket transport}
\label{B}
In the collective basis, the components are obtained as $\tilde{\beta}_k=\sum_j V^*_{kj} \beta_j$. At zero time we can write
\begin{equation}
\tilde{\beta}_k=\beta_0\sum_j V^*_{kj} f_j e^{-i q_0 j}=\frac{\beta_0}{\sqrt{S}}\sum_j e^{i k\theta j} f_j e^{-i q_0 j}=\frac{\beta_0}{\sqrt{S}}\sum_j e^{i(k-k_0)\theta j} f_j,
\end{equation}
which we can estimate in the limit of large $S$ by extending the sum to the whole range of integers (under the assumption that the wavepacket is at all times far from the edges of the chain) and turning it into an integration in the continuum. Notice that for the initial Gaussian distribution
\begin{equation}
\sum_j |f_j|^2 \approx \int_{-\infty}^{\infty} dx \frac{1}{\sqrt{2\pi}w}e^{-\frac{(x-j_0)^2}{2w^2}}=1
\end{equation}
giving the expected normalization condition. We then use know expressions for the Fourier transform of a Gaussian to obtain
\begin{equation}
\sum_j e^{i(k-k_0)\theta j} e^{-\frac{(j-j_0)^2}{4w^2}} \approx \int_{-\infty}^{\infty} dx e^{i(k-k_0)\theta x} e^{-\frac{(x-j_0)^2}{4w^2}} = 2w \sqrt{\pi} e^{i(k-k_0)\theta j_0} e^{-w^2 \theta^2(k-k_0)^2}
\end{equation}
After some simplifications, the expression we will extensively use reads:
\begin{equation}
\sum_j V^*_{kj} f_j e^{-i q_0 j}=\frac{1}{\sqrt{S}}\sum_j e^{i(k-k_0)\theta j} f_j \approx \tilde{f}_k e^{i(k-k_0)\theta j_0}
\end{equation}
where the distribution in k-space is a Gaussian
\begin{equation}
\tilde{f}_k =\frac{1}{\sqrt{\sqrt{2\pi}\tilde{w}}} e^{-\frac{(k-k_0)^2}{4\tilde{w}^2}},
\end{equation}
centered at $k=0$ and with effective width $\tilde{w}=1/(2w\theta)=S/(4\pi w)$. Notice that the validity condition for this discrete to continuum transition is $w \ll S$ reflecting a wide distribution in the k-space $\tilde{w}\gg 1$.\\
To estimate the components in time we use the expression
\begin{equation}
\beta_j(t) = \beta_0 \sum_{k,j'} e^{-i\mathcal{E}_k t-i q_0 j'}e^{-\gamma t/2}V_{jk} V^*_{kj'} f_{j'}=\beta_0 e^{-\gamma t/2} \sum_{k} e^{-i\mathcal{E}_k t}V_{jk} \sum_{k,j'} V^*_{kj'} f_{j'}e^{-i q_0 j'}
\end{equation}
Using the expression from above we simply have
\begin{equation}
\beta_j(t) = \beta_0 e^{-\gamma t/2} \sum_{k} e^{-i\mathcal{E}_k t} V_{jk} \tilde{f}_k e^{i(k-k_0)\theta j_0}=\beta_0 e^{-\gamma t/2} \frac{e^{-i q_0 j}}{\sqrt{S}}\sum_{k} e^{-i\mathcal{E}_k t} e^{-i(k-k_0)\theta j} \tilde{f}_k e^{i(k-k_0)\theta j_0}
\end{equation}
We will use now the Taylor expansion of the energy dispersion relation
\begin{equation}
\mathcal{E}_k \simeq \mathcal{E}_{k_0}-2 \Omega \theta \sin{(k_0 \theta)} (k-k_0)- \Omega \theta^2 \cos{k_0 \theta}(k-k_0)^2 = \mathcal{E}_{k_0}-2 \Omega \theta \sin{q_0} \bar{k}- \Omega \theta^2 \cos{q_0}\bar{k}^2
\label{eq-Omega-k}
\end{equation}
which we re-express by introducing $\bar{k}=k-k_0$ with the new variable centered around $k_0$. Putting together the expression above we can write
\begin{equation}
\beta_j(t) =\beta_0 e^{-i\mathcal{E}_{k_0} t} e^{-\gamma t/2} \frac{e^{-i q_0 j}}{\sqrt{S}}\sum_{\bar{k}} e^{2i\Omega t \bar{k}\theta \sin{q_0}} e^{i \Omega t \theta^2 \cos{q_0}\bar{k}^2} e^{-i\bar{k}\theta (j-j_0)} \tilde{f}_k
\end{equation}
We can now regroup the terms above
\begin{align}
\beta_j(t) &= \beta_0 e^{-i\mathcal{E}_{k_0} t} e^{-\gamma t/2} e^{-i q_0 j} \frac{1}{\sqrt{S}}\frac{1}{\sqrt{\sqrt{2\pi}\tilde{w}}}\sum_{\bar{k}} e^{-i\bar{k}\theta (j-j_0-2\Omega t \sin {q_0})}
                e^{-(\frac{1}{4\tilde{w}^2}+i\Omega t \theta^2\cos {q_0})\bar{k}^2}\\
           &= \beta_0 e^{-i\mathcal{E}_{k_0} t} e^{-\gamma t/2} e^{-i q_0 j} \frac{1}{\sqrt{S}}\frac{1}{\sqrt{\sqrt{2\pi}\tilde{w}}}\sum_{\bar{k}} e^{-i\bar{k}\theta(j-j_0-2\Omega t\sin {q_0} )}
                e^{-\frac{\bar{k}^2}{4\tilde{w}^2}(1+4i\tilde{w}^2\Omega t \theta^2\cos {q_0})}
\end{align}
We peform now the same transition to the continuum and use the following identity
\begin{equation}
\frac{1}{S}\frac{1}{\sqrt{\sqrt{2\pi}\tilde{w}}}\int_{-\infty}^{\infty} dx e^{-i x \theta A}e^{-\frac{x^2}{4\tilde{w}^2}(1+B)}=\frac{2^{3/4}\pi^{1/4}e^{-\frac{A^2 \tilde{w}^2 \theta^2}{1+B}}}{\sqrt{\frac{(1+B) S}{\tilde{w}}}}
\end{equation}
which is valid as long as the real part of $B$ is larger than $-1$. We denoted $A=j-j_0-2\Omega t \sin{q_0}$ and $B=4i\tilde{w}^2\Omega \theta^2\cos {q_0}=i\Omega t \cos{q_0}/w^2$. Working it out the result above reads
\begin{equation}
\beta_j(t)=\beta_0 e^{-i\mathcal{E}_{k_0} t} e^{-\gamma t/2} e^{-i q_0 j}\frac{1}{\sqrt{\sqrt{2\pi}\bar{w}(t)}}e^{-\frac{(j-j_0-2\Omega t \sin{q_0})^2}{4 \bar{w}(t)^2}}
\end{equation}
where the complex time dependent width is
\begin{equation}
\bar{w}(t)=w \sqrt{1+B}= \sqrt{w+i\Omega t \cos{q_0}}=w\sqrt{1+\left(\frac{\Omega t \cos{q_0}}{w}\right)^2}e^{i \phi(t)}
\end{equation}
Notice that the site population is a Gaussian with a time dependent center and variance
\begin{equation}
|\beta_j(t)|=|\beta_0|^2 e^{-\gamma t} \frac{1}{\sqrt{2\pi}\bar{w}(t)}e^{-\frac{(j-\bar{j}(t))^2}{2 |\bar{w}(t)|^2}}.
\end{equation}
\section{Polaritons}
\label{C}
Let us consider the eigenvalue problem of the Tavis-Cummings model plus nearest-neighbor dipole-dipole exchanges. We denote the eigensystem by $\omega_n$ and $\ket{n}$ such that the eigenvalue problem become $\mathcal{H}\ket{n}=\omega_n\ket{n}$ for $n$ running from $1$ to $N+1$. In the single excitation regime the general form of an eigenvector will then be of the form
\begin{equation}
\ket{n}= \sum_{j=1}^N c_j^{(n)}\ket{j}+\beta^{(n)}\ket{1_\text{ph}},
\end{equation}
where normalization requires that $\sum_j |c_j^{(n)}|^2+|\beta^{(n)}|^2=1$. The task is to find all $\omega_n$ and corresponding coefficients of the emitter $c_j^{(n)}$ and photon $\beta^{(n)}$ content in each eigenvector. To this end we write
\begin{equation}
\mathcal{H}\ket{n}=\omega \ket{n}+ \sum_{j=1}^N g_{j} \left(c_j^{(n)}\ket{1_\text{ph}}+\beta^{(n)}\ket{j}\right)+\sum_{j=1}^N \sum_{j'=1}^N \sum_{k}\mathcal{E}_k V_{jk} V_{j'k}^* c^{(n)}_{j'}\ket{j}.
\end{equation}
We have used above the diagonal representation of the dipole-dipole interaction $\mathcal{H}_\text{dd}=2\Omega \sum_k\cos(k\theta)\ket{\tilde{k}}\bra{\tilde{k}}$ and the transformation $\ket{\tilde{k}}=\sum_{k}V_{jk}\ket{j}$ to find the representation $\mathcal{H}_\text{dd}=\sum_k\mathcal{E}_k\sum_{j}\sum_{j'}V_{jk}V^*_{j'k}\ket{j'}\bra{j}$. We will now use the eigenvalue equation $\mathcal{H}\ket{n}=\omega_n\ket{n}$ and consequently $\bra{j}\mathcal{H}\ket{n}=c_j^{(n)}\omega_n$, $\bra{1_\text{ph}}\mathcal{H}\ket{n}=\beta^{(n)}\omega_n$ to find a set of coupled equations
\begin{subequations}
\begin{align}
(\omega_n-\omega)c_j^{(n)}-\sum_{j'=1}^N\sum_k \mathcal{E}_k V_{jk}V_{j'k}^*c_{j'}^{(n)}-g_j \beta^{(n)}=0 \label{eq:pol-1}\\
(\omega_n-\omega)\beta^{(n)}-\sum_{j=1}^N g_j c_{j}^{(n)}=0. \label{eq:pol-2}
\end{align}
\end{subequations}
We can proceed by performing a sum in the upper equation and using that $\sum_{j}g_j c_{j}^{(n)}=(\omega_n-\omega)\beta^{(n)}$ to find
\begin{align}
\left[(\omega_n-\omega)^2 -\sum_{j}g_j^2\right] \beta^{(n)}-\sum_{j'}\sum_k \mathcal{E}_k \left[\sum_{j}g_j V_{jk}\right]V_{j'k}c_{j'}^{(n)} =0.
\end{align}
Notice first that in the absence of particle-particle interactions, the equation above we simply suggest $N-1$ solutions which are degenerate with zero photon components ($\beta^{(n)}=0$) and two non-degenerate polaritonic components with energies $\omega \pm \sqrt{\sum_{j}g_j^2}$.\\
\indent Let us now assume the perfectly symmetric coupling where $g_j=g$. Notice that the sum $\sum_{j} V_{jk}=\sqrt{N}\delta_{k,0}$ selects only the symmetric collective mode with $k=0$, as expected. Also notice that $V_{j'0}=1/\sqrt{N}$ so that one can use again $\sum_{j} c_{j}^{(n)}=(\omega_n-\omega)\beta^{(n)}/g$
\begin{align}
\left[(\omega_n-\omega)^2 -g^2 N\right] \beta^{(n)}-\mathcal{E}_0(\omega_n-\omega)\beta^{(n)} =0.
\end{align}
Again there are $N-1$ degenerate solutions with $\beta^{(n)}=0$ and two polariton energies obtained as solutions of a quadratic equation. Notice that $\mathcal{E}_0=2\Omega$ and the two solutions read
\begin{align}
\omega_{\pm}^\text{sym}=\omega+\Omega\pm \sqrt{g^2 N+\Omega^2}.
\end{align}
Let us now find the eigenvectors corresponding to the polaritonic energies. For symmetric coupling we can rewrite Eq.(\ref{eq:pol-1})
\begin{equation}
(\omega_\pm-\omega) c_j^{\pm}-\frac{\Omega}{N} (c_{j-1}^\pm+c_{j+1}^\pm)-g\beta^{\pm}=0,
\end{equation}
where we have used  $V_{jk}=1/\sqrt{N} e^{-i \theta j k}$, $\mathcal{E}_k= \Omega (e^{i \theta k}+e^{-i \theta k})$. This expression can be satisfied if all the coefficients of the matter part are equal. From Eq.(\ref{eq:pol-2}) and the normalization condition $\sum_j|c_j^\pm|^2+|\beta^{\pm}|^2=1$ we readily find
\begin{equation}
\beta^\pm=\frac{g \sqrt{N}}{\sqrt{(\omega_\pm-\omega)^2+g^2 N}},
\end{equation}
for the photonic part and the following matter contribution
\begin{equation}
c_j^\pm= \frac{\omega_\pm-\omega}{\sqrt{N} \sqrt{(\omega_\pm-\omega)^2+g^2 N}}.
\end{equation}
Notice that, as expected, in the absence of dipole-dipole couplings, the expressions above reduce to the expected $\beta^\pm=1/\sqrt{2}$ and $c_j^\pm=\pm1/\sqrt{2N}$.
For small tunneling rates $\Omega\ll g\sqrt{N}$ we can approximate the polariton energies at $\omega\pm g\sqrt{N}-\Omega$.\\
\indent In the completely asymmetric case where $g_j=g(-1)^j$ we select the asymmetric mode $\sum_{j} (-1)^jV_{jk}\approx \sqrt{N}\delta_{k,N/2}$ (we assumed large even numbers for $N$) such that $\sum_{j} (-1)^j c_{j}^{(n)}=(\omega_n-\omega)\beta^{(n)}/g$ and we can rewrite similarly
\begin{align}
\left[(\omega_n-\omega)^2 -g N^2\right] \beta^{(n)}-\mathcal{E}_{N/2}(\omega_n-\omega)\beta^{(n)} =0.
\end{align}
Again there are $N-1$ degenerate solutions with $\beta^{(n)}=0$ and two polariton energies obtained as solutions of a quadratic equation. Notice that $\mathcal{E}_{N/2}=-2\Omega$ and the two solutions read
\begin{align}
\omega_{\pm}^\text{asym}=\omega-\Omega\pm \sqrt{g^2 N+\Omega^2}.
\end{align}
The polaritons can be found in an analogous manner as in the symmetric case, with the ansatz that the coefficients show the $(-1)^j$ phase dependence.

\end{document}